# Neutron-scattering studies of arsenic sulphide glasses

J.H. Lee, A.C. Hannon* and S.R. Elliott

Department of Chemistry, University of Cambridge, Lensfield Road, Cambridge CB2 lEW, U.K.

*ISIS Facility, Rutherford Appleton Laboratory, Chilton, Didcot OX11 OQX, U.K.

Abstract

High-resolution neutron-scattering measurements have been performed on bulk glasses of $As_2S_3$ and $As_2S_3I_{1.65}$ at a spallation neutron source. For the case of $As_2S_3$, an isotopic-substitution experiment involving the $^{33}S$ isotope has allowed some of the various pair correlations contributing to the second peak in the radial distribution function to be determined by the method of first differences. For the ternary glass, it has been confirmed that iodine bonds preferentially to As atoms, and that on average each As atom is coordinated to one iodine and two sulphur atoms in the first coordination shell.

1. Introduction

Chalcogenide glasses, compounds of Group VI elements (S, Se, Te) with elements such as As, Ge, B etc., have been of great interest for a considerable time, due to their unusual properties (e.g. photo-induced metastability[1]) and many actual and potential technological applications, including low-loss optic fibres, non-linear optical elements etc.[2]. The canonical chalcogenide glass is perhaps the stoichiometric compound arsenic trisulphide, $As_2S_3$, whose crystalline counterpart is orpiment (or auripigment, so-called after its golden-yellow colour). There have been many structural studies of this important glassy material since, in order properly to understand its physical behaviour, a thorough knowledge of its atomic structure is a prerequisite.

A number of diffraction studies of glassy $As_2S_3$ have been performed, using both neutrons[3] and X-rays[4,5] in conventional scattering experiments. Differential anomalous X-ray scattering measurements have also been performed[6] in the vicinity of the As K-edge in order to extract information about the local structure around As atoms. These measurements are closely related to another atom-specific technique that has also been used to probe the local structure in glassy $As_2S_3$, namely X-ray absorption spectroscopy also performed at the As K-edge[7-11]. Other techniques, e.g. Raman scattering[3,12], also shed some light on the local structure of this glass.

The general consensus is that the structure of glassy $As_2S_3$ is for the most part chemically ordered, with each As atom bonded to three S atoms, and each S to two As atoms, in the first coordination shell, as in the structure of the layered crystal, orpiment[13]. The situation regarding the pair-correlation constituents of the <u>second</u> peak in the radial distribution function is less clear. Extended X-ray absorption fine structure (EXAFS) measurements[8-11] indicate that only As-As, and not As-S, correlations contribute (as well as S-S correlations, which cannot be detected by As K-edge EXAFS). However, the differential anomalous X-ray scattering data seem to imply that As-S correlations do contribute to the second coordination shell (as they do also in the orpiment structure[13]).

The incorporation of halogens, e.g. iodine, into glassy arsenic sulphides, produces a dramatic decrease in the glass-transition temperature to below room temperature for the resulting ternary $As_xS_yI_{1-x-y}$ glasses with iodine contents up to 30 at %. The structure of iodine-containing arsenic sulphide glasses, however, is unclear. An early, low-resolution X-ray diffraction experiment[14] concluded that iodine acts as a chain terminator, bonding preferentially to arsenic atoms in place of sulphur, thereby producing a twisted-chain structure comprising
  I
  /
[-S-As-S-] units. Raman scattering spectra, on the other hand, have been interpreted[3,12] as providing evidence for

the existence of discrete $AsI_3$ molecular species dissolved in a glassy As-S matrix. Time-of-flight neutron diffraction data[3] on $(As_2S_3)_{1-x}(AsI_3)_x$ ternary glasses have also been used to support this latter assertion.

In this paper, we report on high-resolution time-of-flight neutron-diffraction experiments carried out on bulk glassy $As_2S_3$ employing, for the first time, isotopic substitution involving the $^{33}S$ isotope in order to obtain additional information on partial atom-atom correlations. In addition, we describe the results of conventional neutron-diffraction experiments on an iodine-containing arsenic sulphide glass, $As_2S_3I_{1.65}$.

2. Experimental details and data analysis

The samples of glassy arsenic sulphide, and the iodine-containing ternary glass, were prepared from the elements, and the isotopically-enriched sample of $As_2S_3$ was prepared from 99 % isotopically-pure[15] $^{33}S$. The constituents in the form of metal pieces, flakes, powder and chips for As, $^{nat}S$, $^{33}S$ and I, respectively, were weighed and placed in silica ampoules that were sealed under vacuum. The charges were then slowly raised in temperature to $800^0C$ in a rocking tube furnace, left at this temperature overnight, and quenched in air to form the glass samples. The mass and atomic densities of the three samples studied are given in Table 1, together with the bound coherent neutron-scattering lengths for the elements and isotopes involved, and the square of compositionally-weighted scattering lengths for the sample compounds.

The composition of the iodine-containing ternary glass chosen for study was $As_2S_3I_{1.65}$ (or equivalently $As_{0.3}S_{0.45}S_{0.25}$), the same as one of the samples studied in an early X-ray diffraction experiment[14]. This particular composition is almost the same as, but slightly sulphur-rich compared with, the member of the family $(As_2S_3)_{1-x}(AsI_3)_x$ having the same atomic fraction of iodine, viz. $As_{0.35}S_{0.4}I_{0.25}$ with x=0.38, and is deep within the glass-forming region of this ternary system[12].

Time-of-flight neutron-diffraction experiments were carried out on the three samples, held at room temperature, using the LAD diffractometer at the ISIS facility, Rutherford Appleton Laboratory. In order to achieve good counting statistics, particularly important for the data subtractions carried out with the isotopically-enriched and natural isotopic abundance $As_2S_3$ glass samples, the measurement runs were very long (44h each for the two $As_2S_3$ samples, 28 h for the $As_2S_3I_{1.65}$ sample, together with 23 h for the empty vanadium can, 15 h for a vanadium rod for normalization purposes and 1 h for the empty spectrometer).

Data were corrected for background, absorption, multiple scattering and inelasticity using the ATLAS suite[16]. After the appropriate corrections for the differential cross section measured at each angle detector bank and subtractions of the self-scattering cross section, the distinct scattering cross sections from all detector banks were obtained and then combined, resulting in the final $i(Q)$. When combining data from different detector banks, particular attention was made to select exactly the same $Q$-range from each angle for both natural and $^{33}S$ isotopically-enriched samples of $As_2S_3$.

The function that is measured in a neutron-scattering experiment is the differential cross-section, given by

$$\frac{d\sigma(Q)}{d\Omega} = i(Q) + I^s(Q), \qquad (1)$$

where $i(Q)$ is the distinct scattering function of interest, resulting from interference of neutrons scattered from different pairs of atoms in the structure, $I^s(Q)$ is the atomic self-scattering function which provides a background signal to $d\sigma/d\Omega$ and which is subtracted to give $i(Q)$, and $Q$ is the momentum transfer (=$(4\pi\sin\theta)/\lambda$, where $2\theta$ is the scattering angle and $\lambda$ is the neutron wavelength).

The distinct scattering function is related to the real-space total correlation function $T(r)$ by means of a Fourier transform:

$$T(r) = T_o(r) + \frac{2}{\pi}\int_0^\infty Qi(Q)M(Q)\sin Qr \, dQ, \qquad (2)$$

where *M(Q)* is a modification function accounting for the fact that the scattering measurements cannot be obtained over an infinite range of momentum transfers as required by the Fourier transform. In this experiment, the maximum value of momentum transfer used was $Q_{max}$=32.6Å$^{-1}$ for all three samples, and the modification function used was the Lorch function[17]. In eqn. (2), the quantity $T_o(r) = 4\pi r \rho_o \left( \sum_i x_i \overline{b_i} \right)^2 \equiv 4\pi r \rho_o \langle b \rangle^2$, where $\rho_0$ is the average atomic density, and $x_i$ and $\overline{b_i}$ are the atomic fraction and coherent scattering length for element i, respectively. The total correlation function is a weighted sum of the partial pair correlation functions $T_{ij}(r)$ for pairs of atoms i and j:

$$T(r) = \sum_{i,j} x_i \overline{b_i} \overline{b_j} T_{ij}(r). \tag{3}$$

3. Results

The distinct scattering functions for the natural isotopic abundance and the isotopically-enriched samples of glassy $As_2S_3$ measured in this study are shown in fig. 1. It can be seen that meaningful oscillations persist up to $Q \sim 30$ Å$^{-1}$, and a value for $Q_{max}$=32.6Å$^{-1}$ was chosen for *all* samples to facilitate comparison between the various data sets. There are large differences evident in *i(Q)* between natural and isotopically-enriched samples, particularly at smaller values of momentum transfer $Q$<15 Å$^{-1}$, as seen in the inset in fig. 1.

There are very pronounced differences in the peak positions and intensities, especially at low *Q*-values, between the iodine-containing and pure (natural abundance) arsenic sulphide glasses (see fig. 2). Note particularly the near-disappearance of the first sharp diffraction peak (FSDP) at $Q \sim 1.2$ Å$^{-1}$. These marked differences are a signature of the structural disruption caused by the incorporation of the chain-terminating iodine atoms into glassy $As_2S_3$.

The real-space correlation functions, $T(r)$, obtained by Fourier transformation of the $i(Q)$ data (eqn. (2)), are shown for the natural and $^{33}S$-enriched samples of $As_2S_3$ in figs. 3 (a,b), respectively, for the range of $r$ encompassing the third peak in $T(r)$. It can be seen that the first peak, corresponding to the first coordination shell, is completely separated from the other peaks in both cases. A measure of how large are the systematic errors in the isotopic-substitution experiments on glassy $As_2S_3$ can be gleaned by comparing the calculated quantity $T(r)/<b>^2$ in each case. From eqn. (2), it can be seen that this quantity is simply related to the *atomic* density, which should be the same in both cases. The comparison is given in fig. 4(a), whence it can be seen that the two curves of $T(r)/<b>^2$ are practically indistinguishable.

The $T(r)$ function obtained for the $As_2S_3I_{1.65}$ glass is shown in fig. 5. It can be seen that the first coordination shell now consists of two clearly resolved peaks. The second peak is also markedly more asymmetric than that characteristic of pure glassy $As_2S_3$.

The structural parameters relating to the first coordination shell can be obtained unambiguously by means of curve fitting, in this case using a peak-shape function involving the Lorch modification function and the $Q_{max}$ value of 32.6 Å$^{-1}$ in the Fourier transformation of the $i(Q)$ data. The results of this peak-fitting procedure are shown in figs. 3(a,b) for the case of the two $As_2S_3$ glass samples, and in fig. 5 for the case of glassy $As_2S_3I_{1.65}$. Values of peak positions and coordination numbers resulting from such peak fits are given in Table 2.

4. Discussion

a) First coordination shell

The results of fitting the first peak in $T(r)$ for the two $As_2S_3$ glasses, given in Table 2, are generally consistent with what is expected for the nature of the first coordination shell in this case, viz, in the case of chemical ordering, a first shell comprising As-S correlations, each As atom being surrounded by $N_{AsS} = 3$ S nearest neighbours and each S atom surrounded by $N_{SAs} = 2$ As nearest neighbours. The value of the As-S bond length found in this study ($r_{AsS} \sim 2.27$ Å) is the same as that found in an earlier time-of-flight neutron diffraction study[3] and close to that ($r_{AsS} = 2.28$ Å) obtained by X-ray diffraction[4,5] and EXAFS[10] experiments.

Although the peak fits to the first peak for the two $As_2S_3$ glasses shown in figs 3(a,b) are generally excellent, there is a small but discernible discrepancy at the base on either side of the peak in both cases. These shoulders to the base of the first peak are revealed as small peaks, located at $r \sim 2$ Å and $\sim 2.5$ Å (see fig. 4(b)), when the $T(r)$ curves, reduced by the factor of $1/<b>^2$, have subtracted from them the fitted first peak (also reduced by the same factor). The fact that these two subsidiary peaks are practically identical for both $As_2S_3$ glasses when plotted in this way (a measure of the atomic-density fluctuation), whereas the spurious termination ripples evident at smaller values of $r$ are <u>not</u> similarly coincident, lends credence to the supposition that these peaks are real and are not artefacts. A small shoulder on the low-$r$ side of the base of the first peak in $T(r)$ has also been seen earlier in a reasonably high-resolution ($Q_{max} = 21.3$ Å$^{-1}$) X-ray diffraction study[5] of glassy $As_2S_3$. Similar small shoulders on either side of the base of the first peak in $T(r)$ have also been observed[18] for the case of $Ge_{25}(As_{1-x}Ga_x)_{10}S_{65}$ glasses measured using high-resolution time-of-flight neutron diffraction.

We propose that the origin of the small subsidiary peaks on either side of the principal first peak in $T(r)$ for glassy $As_2S_3$ lies in *chemical disorder*, i.e. the presence of homopolar S-S and As-As 'wrong' bonds in addition to the heteropolar As-S bonds expected in the case of a perfectly chemically-ordered glass. The glasses examined in this study were quenched from the high temperature of $T \sim 800^0C$. It has been established from EXAFS studies[8] that, with increasing quench temperature in the range 300-800ºC, the nearest-neighbour coordination shell becomes increasingly disordered, as monitored by the static structural contribution to the Debye-Waller factor. The most ordered glass structure was found for samples quenched from $T \sim 300^0C$ (near the crystal melting point). A Raman-scattering study[19] of glassy $As_2S_3$ quenched from various temperatures below $1100^0C$ revealed evidence for a subsidiary peak at ~220 $cm^{-1}$, ascribed to vibrations of As-As bonds, whose intensity increased with quench temperature. (A complementary S-S vibrational band, expected to lie at 450-500 $cm^{-1}$, could not be observed.) However, another Raman-scattering study[12] did find evidence for small peaks at 230 and 490 $cm^{-1}$ in some samples of glassy $As_2S_3$, consistent with the presence of As-As and S-S bonds, respectively.

In crystalline forms of elemental sulphur, the nearest-neighbour S-S bond length is $r_{SS}$ ~2.05Å [20], and thus we identify the small peak in $T(r)$ observed at $r$~2Å with the sulphur-sulphur distance in persulphide bridging units ( $\begin{smallmatrix}\backslash & & /\\ & \text{As-S-S-As} & \\ / & & \backslash\end{smallmatrix}$ ), in agreement with an earlier X-ray study[5]. Likewise, we associate the small peak at $r$ ~2.5 Å with the presence of conjugate As-As bonds, since the As-As bond length in crystalline and amorphous forms of elemental arsenic is[21] $r_{As-As}$ ~2.5 Å. The fact that the two peaks in fig. 4(b) have comparable areas in atomic-number density terms is understandable from the chemical-disorder picture: every sulphur atom removed from an As-S-As unit and inserted into an As-S bond, thereby forming an S-S bond, leaves behind a (reconstructed) As-As bond. The subsidiary shoulders on either side of the base of the first peak of $T(r)$ observed in the case of Ge-As-Ga-S glasses have also been interpreted in terms of homopolar-bond chemical disorder[18].

We turn now to a discussion of the first coordination shell in the structure of glassy $As_2S_3I_{1.65}$ ($As_{0.3}S_{0.45}I_{0.25}$) as revealed by these high-resolution neutron-scattering measurements. As seen in fig. 5, the first coordination shell for this glass clearly consists of two components, and a fit using a peak-shape function as for $As_2S_3$ reveals the existence of two peaks, one located at $r$~2.26 Å and the other at $r$~2.59 Å. A much earlier, very-low-resolution X-ray diffraction study[14], using Cu $K_\alpha$ X-radiation, could not resolve these two constituent peaks but instead found a single broad first peak in the radial distribution function at $r{\sim}2.45$ Å, which is obviously an overlapped combination of the two peaks. A more recent high-resolution neutron-scattering study[3] of the glassy system $(As_2S_3)_{1-x}(AsI_3)_x$ (or equivalently $As_{(2-x)/(5-x)}S_{(3-3x)/(5-x)}I_{3x/(5-x)}$) also reveals a split first coordination shell, with peaks at $r{\sim}2.27$ Å and $r{\sim}2.62$ Å, the latter peak increasing in intensity at the expense of the other with increasing iodine content ($x$). The reduced radial distribution function for the glass in the $(As_2S_3)_{1-x}(AsI_3)_x$ system with the composition closest to that studied here, viz $As_{0.34}S_{0.36}I_{0.3}$ with $x=0.455$, appears to be qualitatively the same as the total correlation function shown in fig. 5.

Note also that, as for the pure $As_2S_3$ glass (figs. 3 and 4), there is a rather pronounced shoulder to the base of the first peak on the low-$r$ side, at $r{\sim}2$ Å. As before, we ascribe this shoulder to the presence of S-S bonds, occurring in this case, not just because of chemical disorder, but also as a result of the stoichiometry.

We associate the peak in $T(r)$ at $r{\sim}2.26$ Å with nearest-neighbour As-S bonds, since the peak position is very close to that found for glassy $As_2S_3$ (see Table 2). We ascribe the second peak in the first coordination shell, at $r{\sim}2.59$ Å, to nearest-neighbour As-I bonds, since the As-I bond length in crystalline $AsI_3$ (consisting of a packing of $AsI_3$ molecules) is[22] $r_{As-I}=2.591$ Å.

The coordination numbers inferred from the areas of these two peaks are given in Table 2. It can be seen that the coordination number for As atoms bonded to I is $N_{IAs}=1.1$, consistent with the idea that indeed iodine acts as a chain terminator, bonding preferentially to arsenic. (Sulphur-iodine bonds have a very low degree of stability and are therefore not expected to be formed[14].)

The coordination numbers for the nearest-neighbour environment of As atoms are $N_{AsS}=2.1$ and $N_{AsI}=0.9$, consistent with the total average coordination number of arsenic being three, as usual. These results can be interpreted in two ways. Either the iodine substitutes randomly for sulphur atoms in bonding to arsenic, thereby forming a twisted chain structure consisting of fragments such as, as proposed by Hopkins et al[14], with arsenic being coordinated to two sulphur atoms and one iodine atom on average. Alternatively, as proposed by Koudelka and Pisarcik[12,13] from a Raman-scattering study, the structure of As-S-I glasses could be composed of discrete $AsI_3$ molecules dissolved in an As-S glass matrix. In this case, the composition of the glass examined in this study, viz. $As_{0.301}S_{0.451}I_{0.248}$, can be rewritten as $(As_{0.218}S_{0.451})(AsI_3)_{0.0827}$. Thus, the overall coordination number for S atoms bonded to As is expected to three, scaled by the proportion of As atoms in the As-S matrix to the total, i..e. $N_{AsS}$ = 218x3/0.301=2.17 and that for I atoms is similarly $N_{AsI}$ = 0.0827x3/0.301=0.82. These are the same as the values found experimentally ($N_{AsS}$ = 2.1, $N_{AsI}$ = 0.9) to within the experimental error.

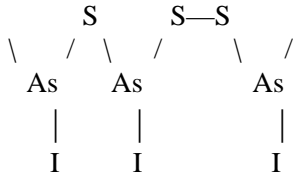

Although analysis of the short-range order (first coordination shell) in $As_2S_3I_{1.65}$ apparently cannot decide between the twisted-chain and $AsI_3$ molecule models for the structural incorporation of iodine, perhaps the $i(Q)$ data at small $Q$, specifically the FSDP, may provide a clue. The near-disappearance of the FSDP for the iodine-containing glass means that the medium-range order that gives rise to this feature, namely quasi-periodic correlations between cation-centred coordination polyhedra (in this case, $AsS_3$ trigonal pyramidal units), must be very nearly completely destroyed in the iodine-containing glass. In the structural model involving discrete $AsI_3$ molecules, there still remains a very sizeable proportion of a highly cross-linked As-S glassy matrix that, presumably, would still give rise to an appreciable FSDP. In the other model, however, where

iodine acts as a chain terminator, [-S-As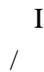-S] chains would be produced, the structural flexibility of which

(associated with chain rotation) would ensure that quasi-periodic As-As correlations would be destroyed. There is thus some evidence that perhaps the random-chain model is the more appropriate.

b) Second coordination shell

The second peak in $T(r)$ for glassy $As_2S_3$ appearing at $r \sim 3.5$ Å (figs. 3(a,b)), will be due to second-neighbour As-As and S-S correlations for the chemically-ordered majority part of the structure (together with second-neighbour As-S correlations associated with the chemically-disordered part). However, there will also be substantial contributions from non-directly-bonded, "interlayer" atomic correlations of all three types if the local structure of the glass is anything like that of the crystal orpiment[13]. In principle, an isotopic-substitution neutron-diffraction study, such as we have performed for this system, should be able to distinguish between the various contributions to the second peak.

The total distinct scattering function, $i(Q)$ (or, equivalently, the total real-space correlation function $T(r)$, related to it by a Fourier transform (eqn. (2)), can be written as a weighted sum of partial atom-atom functions, e.g.

$$i(Q) = \alpha i_{AsAs}(Q) + \beta i_{AsS}(Q) + \gamma i_{SS}(Q) \tag{4}$$

and for isotopic substitution of sulphur:

$$i^*(Q) = \alpha i_{AsAs}(Q) + \beta^* i_{AsS}(Q) + \gamma^* i_{SS}(Q) \tag{5}$$

Thus, the simple first difference of these two quantities is

$$\Delta i(Q) \equiv i^*(Q) - i(Q) = (\beta^* - \beta) i_{AsS}(Q) + (\gamma^* - \gamma) i_{SS}(Q) \tag{6}$$

where the coefficients are given by

$$\alpha = x_{As}^2 b_{As}^2, \tag{7a}$$

$$(\beta^* - \beta) = 2 x_{As} x_S b_{As} (b_S^* - b_S) \tag{7b}$$

and

$$(\gamma^* - \gamma) = x_S^2 (b_S^{*2} - b_S^2). \tag{7c}$$

Thus, this first difference contains structural information only on As-S and S-S pair correlation functions.

The Fourier transform, $\Delta T(r)$, of this first difference is shown in fig. 6(a). It can be seen that the first (As-S) peak has the same shape as in the total $T(r)$ (fig. 3), as expected, but that the second peak is rather different, a pronounced shoulder having appeared on the high-$r$ side. This second peak in the difference spectrum has been fitted with two Gaussian functions (rather than peak-shape functions for ease of fitting), having peak positions at 3.47 and 3.87Å. The dominant contribution to the second peak in the difference function at 3.47 Å will be due to sulphur-sulphur second-neighbour correlations, defining the bond angle subtended at the arsenic atoms, as well as presumably due to non-bonded As-S and S-S correlations with the same range of separations, as found in the orpiment crystal structure[13].

Another first difference can be taken of $i(Q)$ (or $T(r)$) functions for natural-abundance and isotopically-substituted samples that instead involves As-As correlations. Modified functions can be defined as:

$$i_m(Q) = \left(\frac{\gamma^* - \gamma}{\gamma}\right) i(Q)$$

$$= \alpha \left(\frac{\gamma^* - \gamma}{\gamma}\right) i_{AsAs}(Q) + \beta \left(\frac{\gamma^* - \gamma}{\gamma}\right) i_{AsS}(Q) + (\gamma^* - \gamma) i_{SS}(Q) \quad (8a)$$

and

$$i_m^*(Q) = \left(\frac{\gamma^* - \gamma}{\gamma^*}\right) i(Q)$$

$$= \alpha \left(\frac{\gamma^* - \gamma}{\gamma^*}\right) i_{AsAs}(Q) + \beta^* \left(\frac{\gamma^* - \gamma}{\gamma^*}\right) i_{AsS}(Q) + (\gamma^* - \gamma) i_{SS}(Q), \quad (8b)$$

where the asterisk superscript denotes, once more, quantities relating to the isotopically-substituted sample, and the coefficients $\alpha$, $\beta$ and $\gamma$ are as in eqn. (7). Taking the first difference gives

$$\Delta i_m(Q) = i_m(Q) - i_m^*(Q)$$

$$= v i_{AsAs}(Q) + w i_{AsS}(Q) \tag{9}$$

where

$$v = a\left(\frac{\gamma^* - \gamma}{\gamma} - \frac{\gamma^* - \gamma}{\gamma^*}\right) = \frac{(\gamma^* - \gamma)^2}{\gamma \gamma^*} \tag{10a}$$

and

$$w = \beta\left(\frac{\gamma^* - \gamma}{\gamma}\right) - \beta^*\left(\frac{\gamma^* - \gamma}{\gamma^*}\right). \tag{10b}$$

The Fourier transform, $\Delta T_m(r)$, of the difference function $\Delta i_m(Q)$ is shown in fig. 6(b). As for the other difference function, $\Delta T(r)$, shown in fig. 6(a), the first peak is due to As-S correlations, but the shoulder at $r \sim 2.5$ Å on the high-$r$ side of the base of the first peak in $\Delta T_m(r)$ is more pronounced than in fig. 6(a), lending support to the assertion made earlier that it is due to As-As 'wrong' bonds. The second peak in $\Delta T_m(r)$ is at almost exactly the same position as that in $\Delta T(r)$, but the pronounced shoulder on the high-$r$ side in $\Delta T(r)$, evident in fig. 6(a), is missing in $\Delta T_m(r)$, leaving only a small peak separated from the main second peak. This second peak in $\Delta T_m(r)$ can be fitted by three Gaussian peaks, as shown in fig. 6(b), positioned at 3.16, 3.50 and 3.98 Å. However, the position of the component peak at highest $r$ is somewhat imprecise; variations in the upper cut-off for fitting produced variations in the peak position of $\Delta r \sim \pm 0.1$ Å. The small first peak at $r \sim 3.16$ Å can possibly be ascribed to As-As 'interlayer' non-bonded correlations; such atomic correlations occur in orpiment at[13] 3.19 Å. The main contribution at 3.50 Å is practically the same as for $\Delta T(r)$ (3.47 Å), and this similarity reflects the fact that, at least in the orpiment crystal structure, the average bond angles subtended at sulphur and at arsenic atoms are very similar (97.5° and 99°, respectively), and hence the average second-neighbour As-As and S-S distances should be almost indistinguishable.

It is a little disappointing that isotopic substitution has not been successful in differentiating all the various second-neighbour contributions to the second peak in $T(r)$ for glassy $As_2S_3$ and hence cannot be used to determine separately the bond angles subtended at As and S atoms. However, the fact that the pronounced shoulder observed on the high-$r$ side of the second peak in $\Delta T(r)$ is revealed as a small separate peak in $\Delta T_m(r)$, means that this feature, also responsible for the high-$r$ asymmetry of the second peak in $T(r)$ (e.g. see fig. 4), is probably due primarily to non-bonded As-S correlations, since it appears in both difference functions.

We turn now to an examination of the second main peak in $T(r)$ for the $As_2S_3I_{1.65}$ glass (see fig. 5). Comparison with fig. 3(a), showing $T(r)$ for the natural-abundance $As_2S_3$ glass sample, shows that the peak maximum has shifted to $r \sim 3.65$ Å and that the high-$r$ side of the peak has become markedly asymmetric. Similar behaviour was found by Kameda et al[3] in a neutron-diffraction study of a series of $(As_2S_3)_{1-x}(AsI_3)_x$ glasses with $0<x<0.65$, who found evidence for the growth of an additional peak at 3.98 Å with increasing iodine content. The I-I second-neighbour distance in crystalline $AsI_3$ is $r=3.959$ Å [22] and so it is tempting to associate the high-$r$ asymmetry of the second peak in $T(r)$ for the As-S-I glass with the formation of discrete $AsI_3$ molecules dissolved in a glassy As-S matrix. However, Hopkins et al[14] have also ascribed the appearance of correlations at r $\sim$3.9 Å to I-I distances in a twisted chain model where I atoms, bonded to different As atoms, are separated by an As-S-S-As chain. Diffraction data alone are not really capable of distinguishing between these two alternative structural models.

## 5. Conclusions

A high-resolution pulsed neutron-diffraction study has been performed to investigate the atomic structure of glassy $As_2S_3$ and $As_2S_3I_{1.65}$ ($As_{0.3}S_{0.45}I_{0.25}$). In addition, an isotopic-substitution experiment using the $^{33}S$ isotope has been carried out for glassy $As_2S_3$. Even with isotopic substitution, the main pair correlations to the second peak in the radial distribution function for glassy $As_2S_3$ could not be resolved, and hence the sulphur and arsenic bond angles could not separately be determined but they must be very similar in value. However, the high-$r$ asymmetry in the second peak can be ascribed probably to the contribution of non-bonded As-S correlations occurring at r ~3.9 Å using this method. Small shoulders to the low- and high-$r$ sides of the base of the first (As-S) peak in the radial distribution function are ascribed to S-S and As-As 'wrong' bonds, respectively, a manifestation of the chemical disorder characteristic of this glassy sample quenched from a high temperature ($800^0C$). A recent density-functional-based tight-binding molecular-dynamics simulation[24] of a-$As_2S_3$, produced results very similar to those reported here. In particular, the small shoulders lying on the low-$r$ and high-$r$ sides of the first As-S peak in the RDF were definitely identified as being due to S-S and As-As bonds respectively, in a model containing such homopolar bonds.

In the As-S-I glass, the first coordination shell has been found to be clearly split into two components, one due to As-S and one due to As-I nearest-neighbour contributions. A marked high-$r$ asymmetry of the second peak in the radial distribution function was found for the I-containing glass compared to pure glassy $As_2S_3$ due to the presence of I-I correlations. However, analysis of these diffraction data to give short-range structural information is alone incapable of being used to distinguish between two competing structural models for the incorporation of iodine into glassy arsenic sulphide, namely the formation of discrete $AsI_3$ molecules dissolved in an As-S glassy matrix, or alternatively the formation of $\genfrac{}{}{0pt}{}{I}{[-S-As-S-]}$ chains with the iodine atoms acting as chain terminators. However, the fact that the intense first sharp diffraction peak (FSDP) characteristic of glassy $As_2S_3$ is practically absent in the iodine-containing glass perhaps favours the random-chain model, where there is no cross linking and hence little structural frustration forcing the creation of the medium-range order required to produce the FSDP.

**References**

[1] K. Shimakawa, A. Kolobov and S.R. Elliott, Adv. Phys. **44,** 475 (1995).

[2] A. Zakery and S.R. Elliott, J. Non-Cryst. Sol. **330**, 1 (2003).

[3] Y. Kameda, Y. Sugawara and O. Uemura, J. Non-Cryst. Sol. **156-158,** 725 (1993).

[4] A.J. Leadbetter and A.J. Apling, J. Non-Cryst. Sol. **15,** 250 (1974).

[5] A.J. Apling, A.J. Leadbetter and AC. Wright, J. Non-Cryst. Sol. **23,** 369 (1977).

[6] Q. Ma, W. Zhou, D.E. Sayers and M.A. Paesler, Phys. Rev. **B 52,** 10025 (1995).

[7] A.J. Lowe, G.N. Greaves and S.R. Elliott, Phil. Mag. **B 54,** 483 (1986).

[8] C.Y. Yang, D.E. Sayers and M.A. Paesler, Phys. Rev. **B 36,** 8122 (1987).

[9] C.Y. Yang, M.A. Paesler and D.E. Sayers, Phys. Rev. **B 36,** 9160 (1987).

[10] C.Y. Yang, M.A. Paesler and D.E. Sayers, Phys. Rev. **B 39,** 10342 (1989).

[11] G. Pfeiffer, J.J. Rehr and D.E. Sayers, Phys. Rev. **B 51,** 804 (1995).

[12] L. Koudelka and M. Pisarcik, J. Non-Cryst. Sol. **64,** 87 (1984).

[13] D.J.E. Mullen and W. Nowacki, Zeit. Krist. **136,** 48 (1972).

[14] T.E. Hopkins, R.A. Pasternak, E.S. Gould and J.R. Herndon, J. Phys. Chem. **66,** 733 (1962).

[15] The $^{33}$S isotope was purchased from JSC JV Isoflex, Schukinskaya Street 12-1, 123182 Moscow, Russia, and had a reported isotopic composition of 99.15 % $^{33}$S, 0.64% $^{34}$S and 0.21 % $^{32}$S.

[16] A.C. Hannon, W.S. Howells and A.K. Soper, IOP Conf. Ser. **107,** 193 (1990).

[17] E. Lorch, J. Phys. **C2,** 229 (1969).

[18] A.C. Hannon and B.G. Aitken, J. Non-Cryst. Sol. **256-257**, 73 (1999).

[19] K. Tanaka, S. Gohda and A. Odajima, Solid State Comm. **56,** 899 (1985).

[20] N.A. Greenwood and A. Earnshaw, "Chemistry of the Elements" (Pergamon, Oxford: 1984), p. 774.

[21] G.N. Greaves, S.R. Elliott and E.A. Davis, Adv. Phys. **28,** 49 (1979).

[22] R. Enjalbert and J. Galy, Acta Cryst. **B 36,** 914 (1980).

[23] L. Koudelka and M. Pisarcik, Solid State Comm. **41,** 115 (1982).

[24] S.I. Simdyankin, S.R. Elliott, Z. Hajnal, T.A. Niehaus and Th. Frauenheim, Phys. Rev. **B,** (To be published) (cond-mat/0312224).

Table 1. Sample details

| Element/isotope | Scattering length, $b$ (fm) | | |
|---|---|---|---|
| $^{nat}$S | 2.847 | | |
| $^{33}$S (99.15%) | 4.73 | | |
| As | 6.58 | | |
| I | 5.28 | | |
| Sample | Mass density (g cm$^{-3}$) | Atomic density (Å$^{-3}$) | $<b>^2$ (barns) |
| As$_2$$^{nat}$S$_3$ | 3.1679* | 0.03877 | 0.188 |
| As$_2$$^{33}$S$_3$ | 3.2033† | 0.03877 | 0.299 |
| As$_2$S$_3$I$_{1.65}$ | 3.5812$^x$ | 0.03149 | 0.209 |

*Measured using a micropycnometer at ISIS; †calculated assuming that the atomic densities of both As$_2$S$_3$ samples are identical; $^x$measured by flotation in acetone.

Table 2. Values of average bond length and coordination number for the first coordination shell found from peak fitting.

| Sample | $r_{AsS}$ (Å) | $N_{AsS}$ | $N_{SAs}$ | | |
|---|---|---|---|---|---|
| $As_2{}^{nat}S_3$ | 2.27 | 2.8 | 1.8 | | |
| $As_2{}^{33}S_3$ | 2.27 | 2.8 | 1.8 | | |
| Sample | $r_{AsS}$ (Å) | $r_{AsI}$ (Å) | $N_{AsS}$ | $N_{AsI}$ | $N_{IAs}$ |
| $As_2S_3I_{1.65}$ | 2.26 | 2.59 | 2.1 | 0.9 | 1.1 |

**Figure Captions**

Fig. 1. The distinct scattering function, *i(Q)*, obtained from time-of-flight neutron diffraction for samples of bulk glassy $As_2S_3$ containing sulphur with the natural isotopic abundance (dashed curve) and 99 % enriched in the $^{33}S$ isotope (solid curve). The inset shows a magnification of the low-$Q$ region.

Fig. 2. The distinct scattering function, *i(Q)*, obtained from time-of-flight neutron diffraction for bulk glassy $As_2S_3I_{1.65}$ (solid curve) compared with that for bulk glassy $As_2S_3$ with the natural isotopic abundance of sulphur. The inset shows a magnification of the low-$Q$ region.

Fig. 3. The total correlation function, *T(r)*, obtained by Fourier transformation of *i(Q)* (eqn. (2)) for bulk glassy $As_2S_3$ containing: a) the natural isotopic abundance of sulphur; and b) 99 % enriched in the $^{33}S$ isotope. In both cases, the peak-shape function fit to the first peak is shown as the dashed curve.

Fig. 4. The total correlation function, *T(r)*, normalized by the quantity $<b>^2$ for glassy $As_2S_3$ for: a) natural isotopic abundance of sulphur (dashed curve) and 99 % enriched in the $^{33}S$ isotope (solid curve); b) the same as a), but with the first peak subtracted using the fit shown in figs. 3 (a,b).

Fig. 5. The total correlation function, *T(r)*, for bulk glassy $As_2S_3I_{1.65}$ obtained by Fourier transformation of *i(Q)* (solid curve), with the peak-shape function fit to the split first coordination shell shown by the dashed curve.

Fig. 6. The first difference for the total correlation function for bulk glassy $As_2S_3$ obtained from the data for the natural isotopic abundance sample and the $^{33}S$-enriched sample: a) $\Delta T(r)$ (c.f eqn. (6)); b) $\Delta T_m(r)$ (c.f eqn. (9)). In both cases, the fit made to the second peak using three Gaussian functions is shown by the dashed-dotted curve, with the three individual peaks shown as the dashed curves.

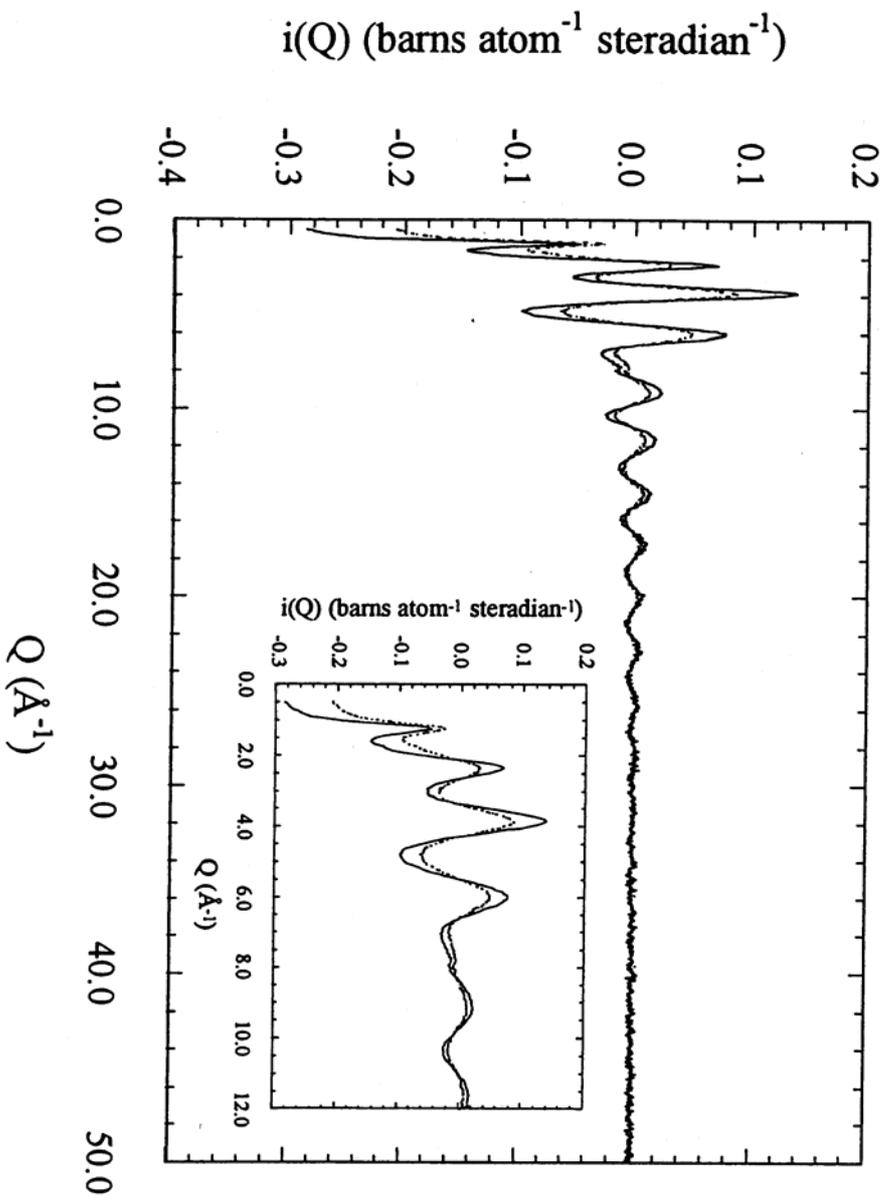

fig 1

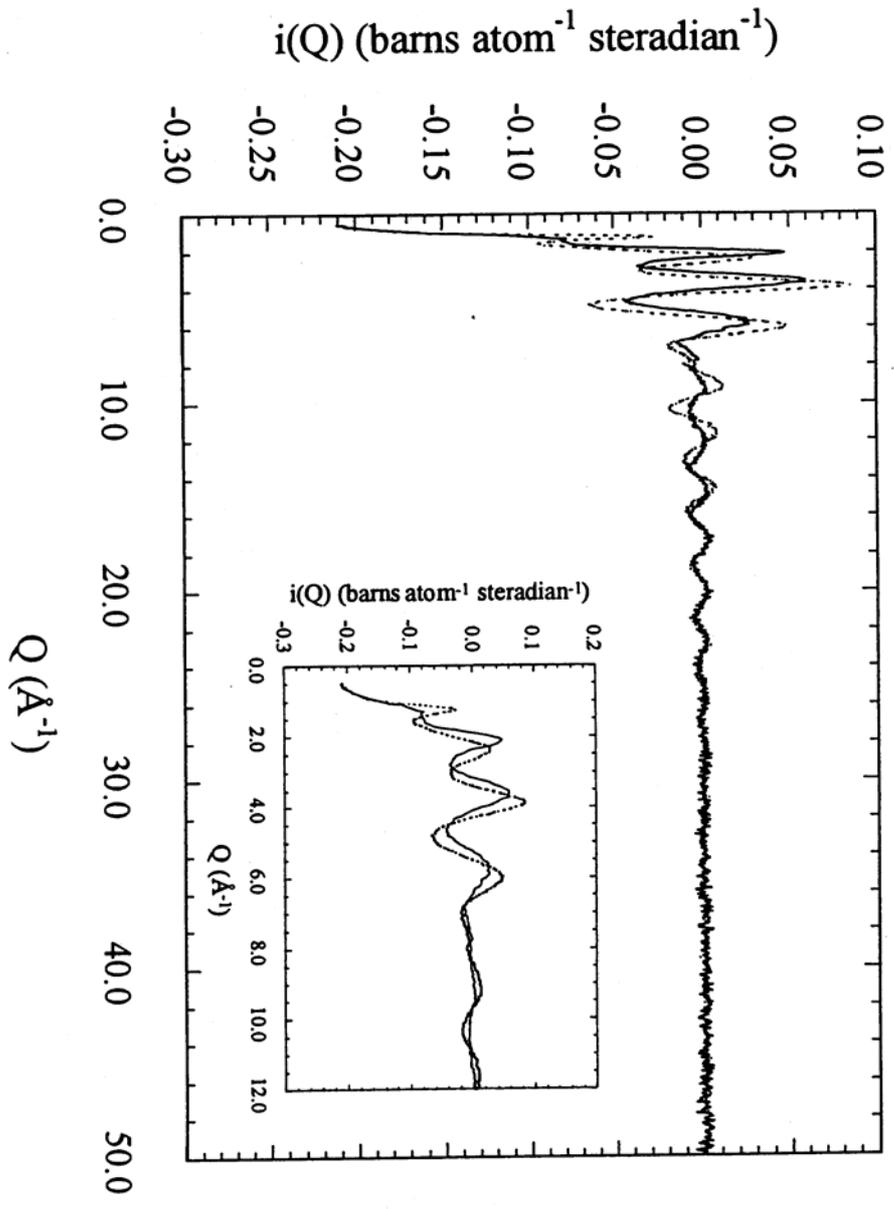

Fig 2

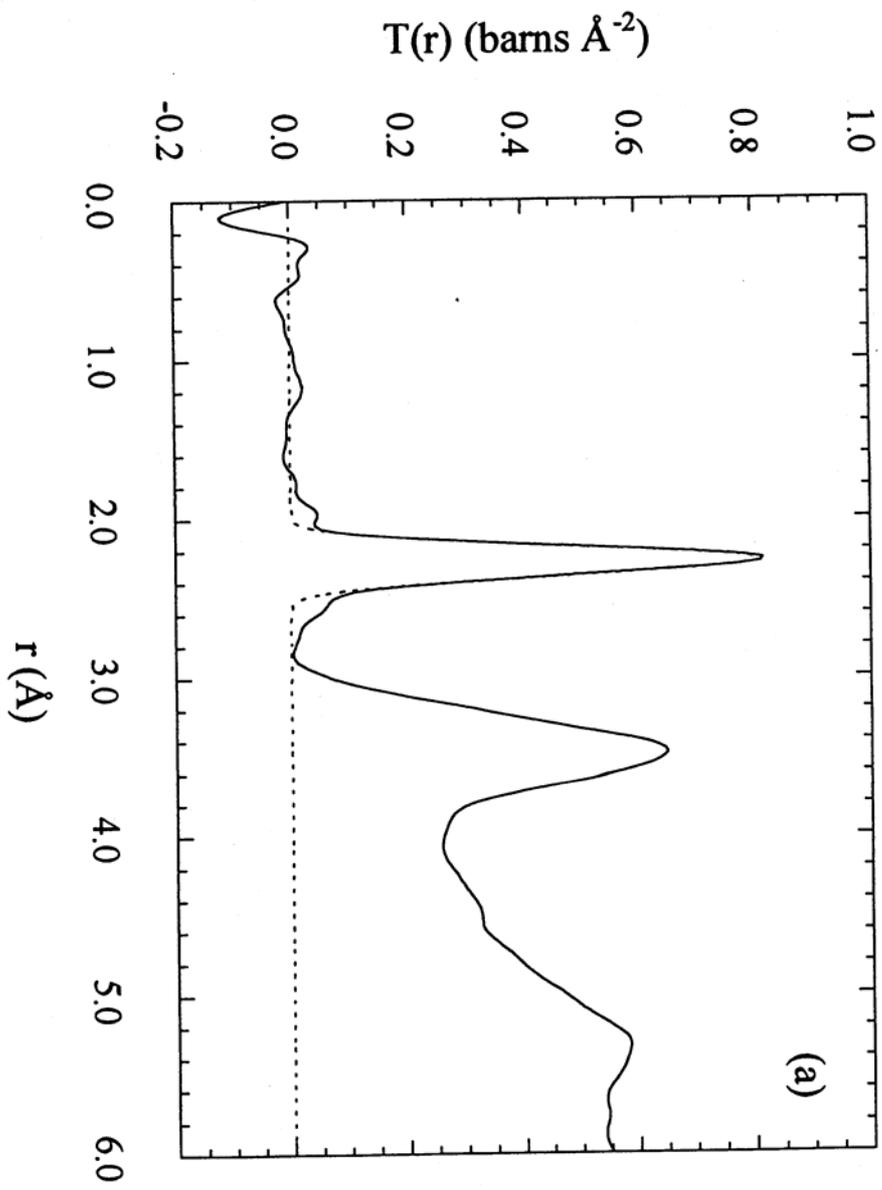

Fig 3 (a)

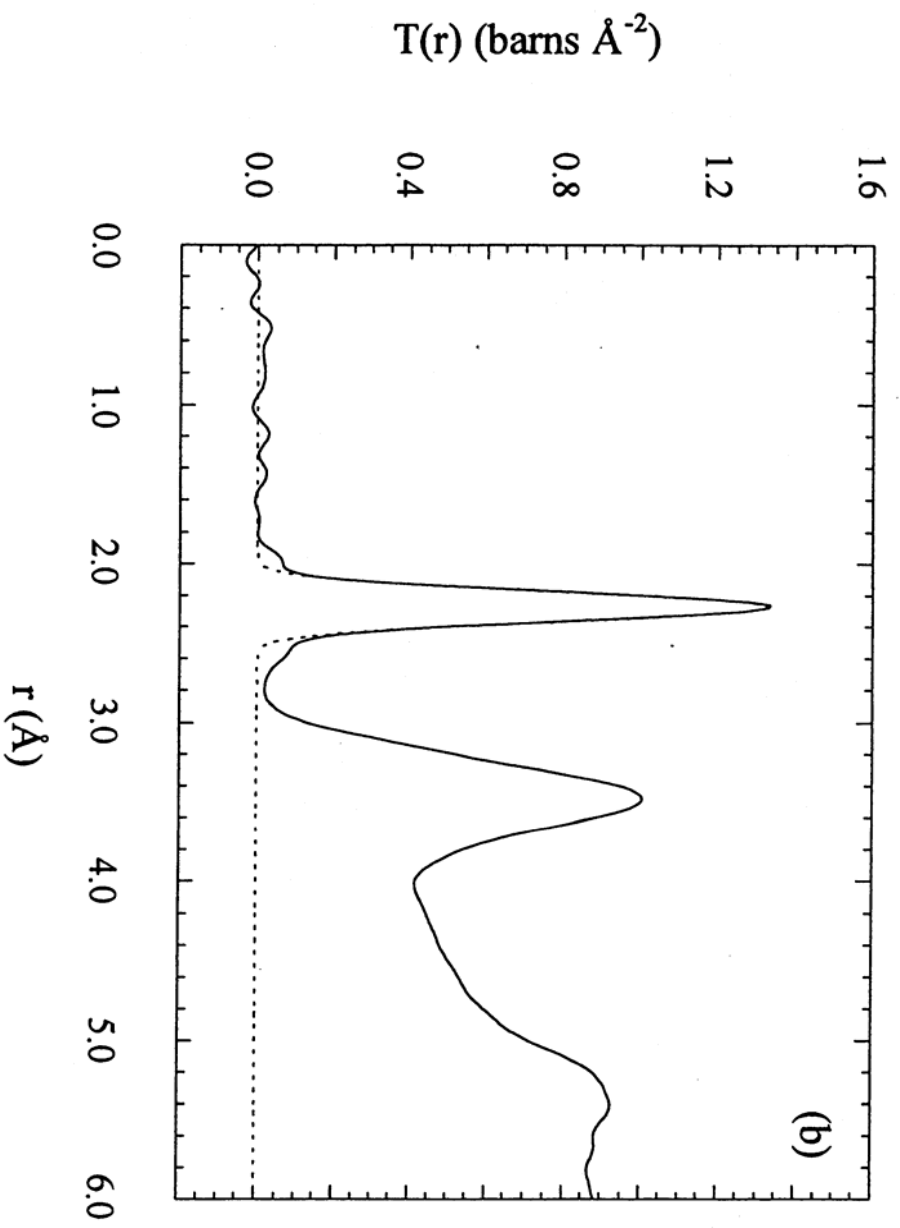

Fig 3 (b)

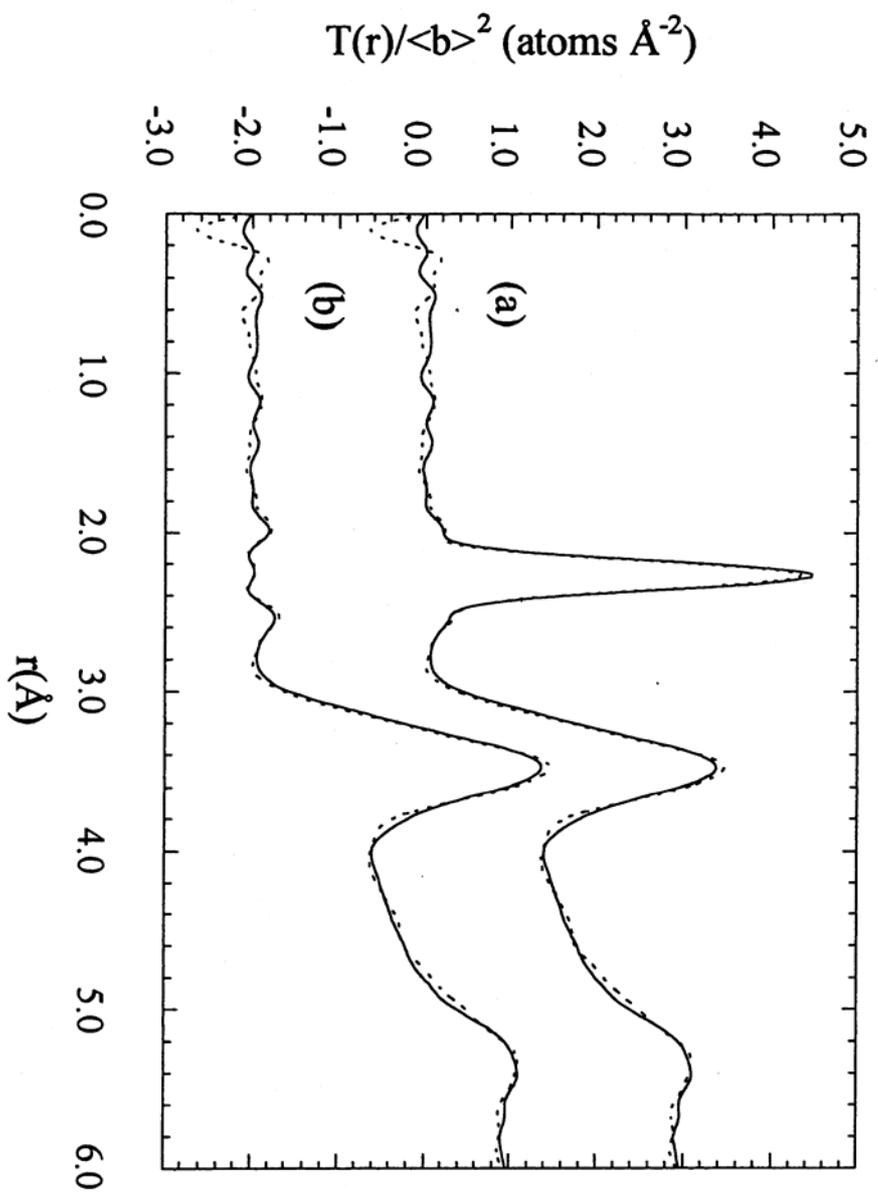

fig 4

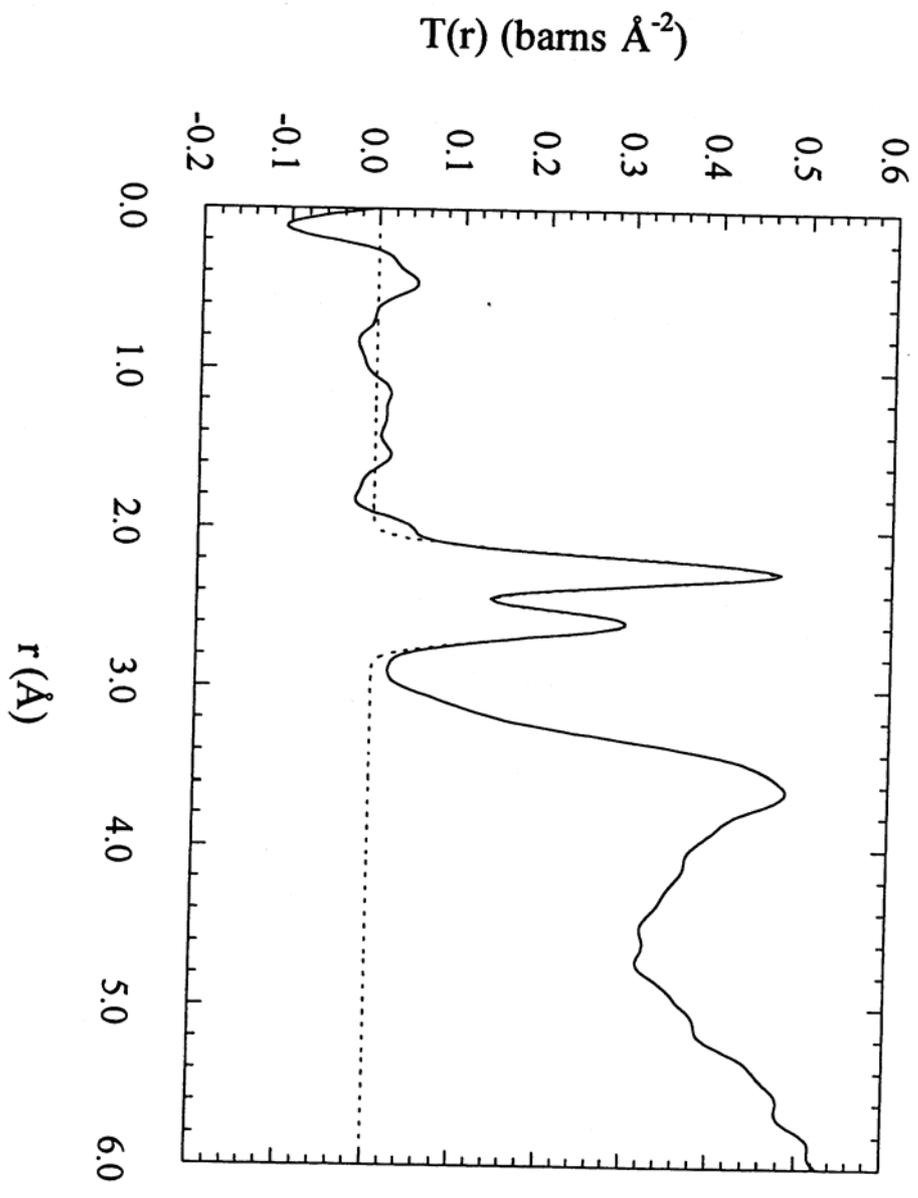

Fig 5

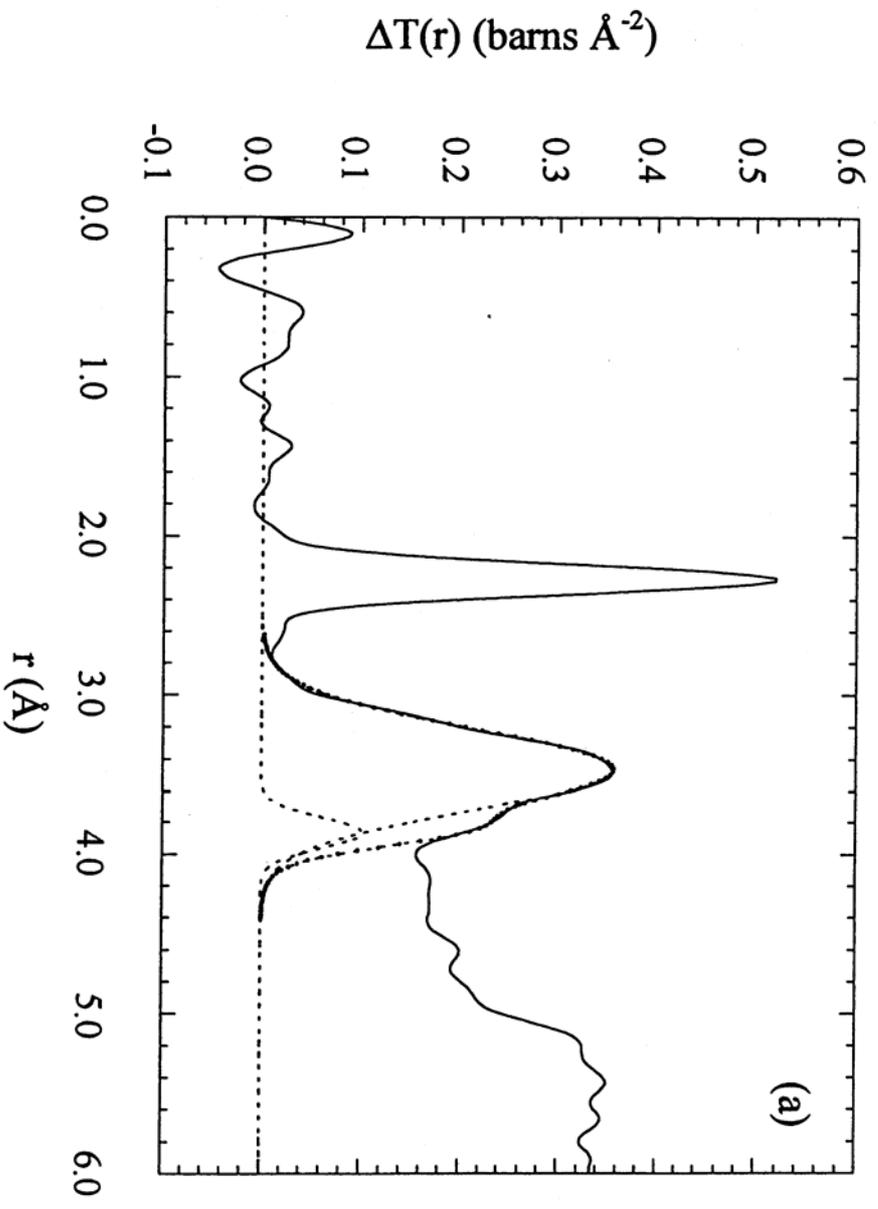

Fig 6 (a)

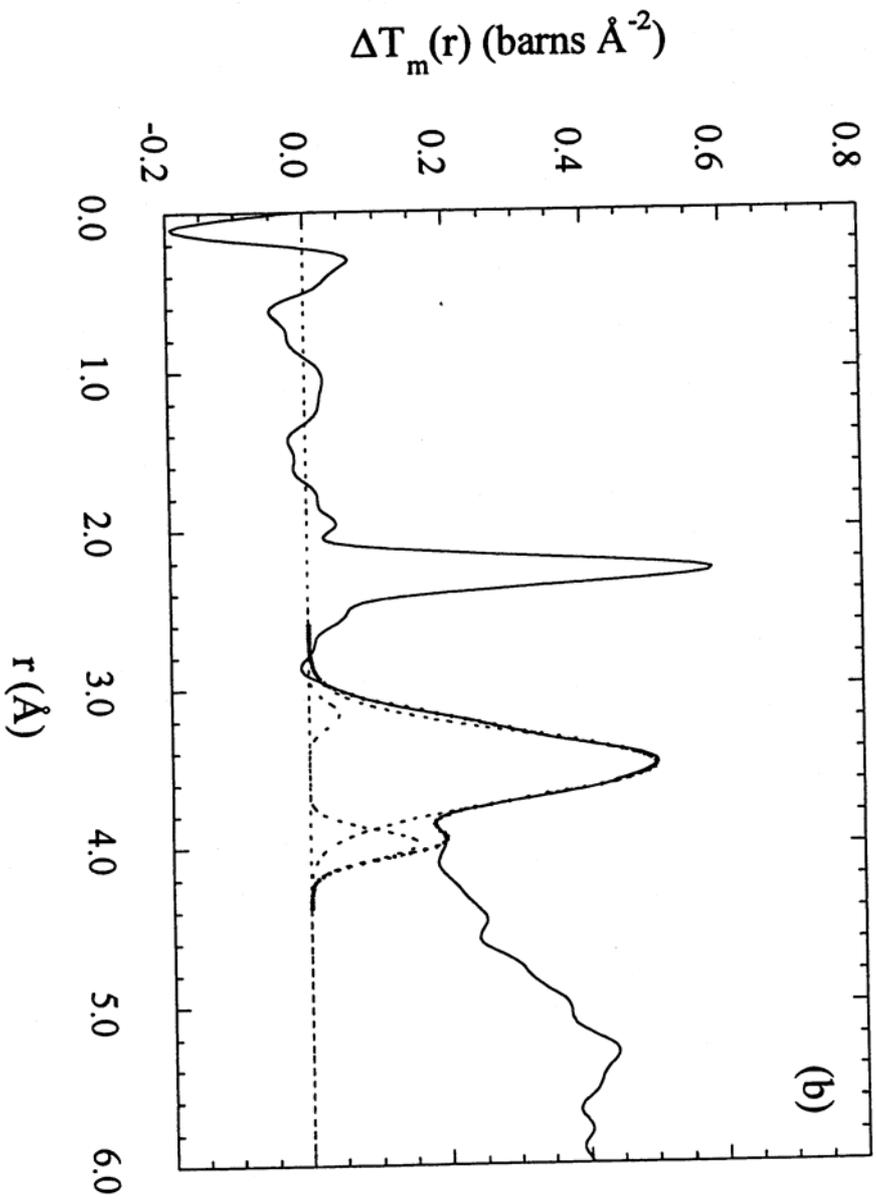

Fig 6 (b)